\documentclass{IEEEtran}
\usepackage[utf8]{inputenc}
\usepackage{amsmath}

\usepackage{csquotes}
\usepackage[backend=biber,style=numeric,sorting=none,natbib=true]{biblatex}
\usepackage[colorlinks]{hyperref}

\usepackage{multirow}
\usepackage{booktabs}
\usepackage{threeparttable}

\usepackage{pifont}
\newcommand{\cmark}{\ding{51}}

\usepackage[binary-units=true]{siunitx}

\usepackage{varioref}

\usepackage{balance}

\title{Improving security for users of decentralized exchanges through multiparty computation}

\author{\IEEEauthorblockN{Robert~Annessi\IEEEauthorrefmark{1} and Ethan~Fast\IEEEauthorrefmark{2}}\\
\IEEEauthorblockA{Nash Exchange}\\
Email: \IEEEauthorrefmark{1}robert@nash.io, \IEEEauthorrefmark{2}ethan@nash.io}

\begin{document}
\maketitle

\begin{abstract}
Decentralized cryptocurrency exchanges offer compelling security benefits over centralized exchanges: users control their funds and avoid the risk of an exchange hack or malicious operator. However, because user assets are fully accessible by a secret key, decentralized exchanges pose significant internal security risks for trading firms and automated trading systems, where a compromised system can result in total loss of funds. Centralized exchanges mitigate this risk through API key based security policies that allow professional users to give individual traders or automated systems specific and customizable access rights such as trading or withdrawal limits. Such policies, however, are not compatible with decentralized exchanges, where all exchange operations require a signature generated by the owner's secret key. This paper introduces a protocol based upon multiparty computation that allows for the creation of API keys and security policies that can be applied to any existing decentralized exchange. Our protocol works with both ECDSA and EdDSA signature schemes and prioritizes efficient computation and communication. We have deployed this protocol on Nash exchange, as well as around several Ethereum-based automated market maker smart contracts, where it secures the trading accounts and wallets of thousands of users. 
\end{abstract}

\section{Introduction}
Centralized cryptocurrency exchanges hold custody of user funds and promise access to those funds when requested. As these exchanges often own large amounts of cryptocurrency funds in aggregate across their users, they are very attractive targets for criminals. Hundreds of millions of dollars of cryptocurrency have been lost over the years through hacks or malicious operators. These security problems are a major driver behind the success of decentralized\footnote{In this paper, we use the established term \textit{decentralized}, but we could equally use the terms \textit{non-custodial} or \textit{self-custodial}.} exchanges, where users hold their own secret keys.

Decentralized exchanges must overcome their own set of security challenges, however. In particular, automated trading algorithms deployed on decentralized exchanges must be trusted to hold and secure a secret key which fully controls user funds. Such trading algorithms are often hosted on cloud infrastructure and operate on shared accounts within trading firms, posing social and technical security risks. This is a significant limitation for professional users such as market makers. These users cannot provide restricted access to their account's capabilities such that their trading software may only trade; their software could as well withdraw funds, which significantly increases downside risk in the event their software or systems are compromised. 

Centralized exchanges overcome such limitations by offering security policies, usually based on API keys, that include limiting the capabilities of trading algorithms to withdraw funds or allowing accounts to trade funds only within certain limits. This functionality is straightforward in centralized settings because the exchange can simply create random identifiers and passwords to enforce security policies via hosted software. While these policies are only as strong as the security of the centralized exchange itself, they do reduce the downstream damage if a user's automated trading infrastructure is compromised. Unfortunately, such security policies do not exist on most decentralized cryptocurrency exchanges today due to the fact that all operations are governed by digital signatures produced via a single secret key. 

In this paper, we introduce a novel protocol that allows for the creation of API key based security policies for any decentralized cryptocurrency exchange. The protocol we describe is currently deployed on Nash~\cite{nash}, but can be applied to any smart contract based exchange protocol. In fact, the Nash mobile wallet also applies this protocol to Uniswap~\cite{uniswap} and 1inch~\cite{1inch}, allowing users to trade over these contracts without ever accessing their full secret key. The protocol we introduce has several important properties:

\begin{itemize}
    \item Policies can be applied and extended to the signature schemes used by the most popular blockchains: in particular ECDSA and EdDSA and the most common elliptic curves on which they are used.
    \item There are no limitations on the kinds of security policies which can be enforced. Exchanges can create policies for withdrawal limits, trading limits, delayed withdrawals, or, for example, other more esoteric functionality such as geolocation or biometric information.
    \item Users fully control their assets and the secret keys that govern them. Secret keys are never accessible by any other system or party.
    \item Signatures produced via the protocol presented in this paper are efficient in computation, rounds of communication, and bandwidth; this is important to professional traders who require low latency for fast order execution and may trade many times per second.
\end{itemize}

While trading applications are the primary focus of this paper, the protocol we describe can easily be extended to enhance user wallet security as well.

\section{Background}
\subsection{Digital Signature Schemes}
A digital signature scheme consist of three algorithms~\cite{katz}: a key generation algorithm, a signing algorithm, and a verification algorithm. The key generation algorithm generates a secret signing key and a public verification key. The signing algorithm is used to generate a signature for a particular message using the secret key. The verification algorithm is used with the public key in order to verify the signature for a particular message (and therefore confirm its authenticity).

\subsection{ECDSA}
\label{sec:ecdsa}
An ECDSA signature consists of a tuple of integers $(r, s)$ that is computed as follows~\cite{JohnsonEllipticCurveDigital2001}:
\begin{enumerate}
	\item A cryptographically secure nonce $k$ is generated with $0 < k < n$, with $n$ the order of the elliptic curve. The implementer must ensure that the nonce $k$ cannot be guessed and is not reused for any two messages. Otherwise, anyone who can guess the nonce used in one signature or who observes two signatures where the identical nonces were used twice can derive the secret key.
	Since there have been a quite a number of insecure implementations, a standard was created on how to generate cryptographically secure nonces deterministically, which is now used in most implementations~\cite{rfc6979}.

	\item A point on the elliptic curve is computed as $(x, y) = k \cdot G$, with $G$ being the generator point of the curve.
	The $x$ coordinate of that point is used as the first part of the signature, i.e., $r = x$.

	\item The second part of the signature $s$ is computed from the secret key $d$, the hash of the message $z$, the first part of the signature $r$, and the nonce $k$ as follows: $s \equiv k^{-1} \cdot (z + r \cdot d) \mod n$.
\end{enumerate}

\subsection{EdDSA}
Signature generation in EdDSA works similar to ECDSA.
An EdDSA signature also consists of a tuple of integers $(r, s)$, but computation differs slightly:
\begin{enumerate}
    \item First, the secret key is hashed. The first half of the hash is used as signing key $a$ and the second half is used in nonce generation.
    \item A cryptographically secure nonce is also required in EdDSA, but generating it is not left to the implementer. Instead, the nonce $k$ is computed by hashing the concatenation of the nonce-part generated in the first step with the message to be signed.
    \item The first part of the signature ($r$) is computed in exactly the same way as in ECDSA (besides the fact that there is a different elliptic curve involved).
    \item The second part of the signature $s$ is computed as $s = (k + H(r | A | M) \cdot a) \mod L$, where $A$ is the public key, $M$ the message, and $L$ the order of the Edwards curve. Note that in contrast to ECDSA, an EdDSA signature does not involve computing the inverse of the nonce $k$.
\end{enumerate}

\subsection{Threshold Signature Schemes}
Threshold signature schemes are an application of secure multiparty computation.
The basic idea in threshold signature schemes is that multiple parties are required to generate a signature. For this purpose, the secret key is split into multiple \textit{key shares}, one for each party. The parties use their key shares to compute partial results, which are then combined to produce the signature. The resulting signature is indistinguishable from a signature computed by a single party with the secret key, such that only the signing and key generation algorithms are replaced in threshold signature schemes but the verification algorithm remains unchanged.

A threshold $t < n$ is defined such that any subset of $t$ or more parties can jointly generate signatures, but any subset smaller than $t$ can not.
It is important to note that, with threshold signature schemes, the actual secret key never appears during signature generation.

Efficient threshold signature schemes have existed for some signature schemes (e.g., Schnorr~\cite{schnorr}) for quite some time, but have only more recently been proposed for ECDSA. Designing an efficient ECDSA threshold signature scheme is challenging because signature generation involves the computation of the inverse of the nonce $k$, but the threshold setting dictates that no single party must know the nonce (as it could otherwise derive the secret key from the resulting signature). EdDSA, on the other hand, does not involve the computation of the inverse of the nonce so that designing a threshold EdDSA signature scheme is significantly less of a challenge.

\section{Design Goals and Protocol Abstraction}
Our API key protocol has been developed to account for three major design goals. First, it must be non-custodial, such that a third party never has control over user funds. Second, it must allow for the creation of flexible security policies (e.g., withdrawal limits based on time windows or addresses, market volume restrictions, etc.). Finally, it must minimize communication overhead, in terms of both bandwidth and rounds of communication for interacting parties.

The importance of the first design goal is self-evident, as the main benefit of interacting with a decentralized exchange is self-custody. A protocol for decentralized API keys should maintain that property. The second goal seeks to accommodate the widest possible variety of security policies; an ideal protocol would allow for the ability to construct any computable policy. The third goal is important as it allows the protocol to serve the needs of professional traders. Traders interacting with centralized exchanges place orders many times per second.

The protocol we have designed works broadly as follows (see technical details in Section~\ref{sec:protocol}): Instead of generating signatures with a secret key residing just on a single machine, we employ a threshold signature scheme with two parties: the client and the exchange. Both parties need to cooperate in order to generate signatures. While the user is still in possession of the secret key, it is never actively used for trading. With the secret key, the user only creates API keys along with policies that describe the capabilities of these API keys, i.e., under what circumstances the exchange should cooperate with the client on generating a signature. A policy can describe arbitrary properties. For example, a policy can restrict an API key to trade on certain markets only and to withdraw funds only to a specific address. So the secret key can be kept on a trusted, potentially even air gapped, device while API keys are used for day-to-day activities such as trading.

Theoretically, multi-signatures could also be used, but threshold signature schemes provide more flexibility than multi-signatures, because the underlying blockchains are not required to provide support specifically.
Contrary to multi-signatures, the signature resulting from a threshold signature scheme is indistinguishable from a conventional ECDSA signing algorithm.
For this reason, the verification algorithm can remain unchanged, and even the key generation algorithm can remain unchanged.
Furthermore, it is impossible to secure an existing address by multi-signatures.
A new address needs to be generated and funds transferred to that new address.
An address can be secured with threshold signatures, however, without generating a new address and moving funds.

Unfortunately, threshold ECDSA schemes are challenging. We review existing threshold ECDSA schemes for their suitability to fit our protocol's design goals in Section~\ref{sec:evaluation}. One major difference in our protocol is that the user knows their secret key. For this reason, API keys can be generated by the user offline on a trusted device, and the generated API key can then be transferred to some untrusted device without harming security. Furthermore, we aim to reduce the latency as perceived by clients as much as possible. We achieve this by shifting most of the computation and communication to earlier points in time, before actual trades are conducted. In this way, a client can conduct a trade within milliseconds (sub-millisecond even for EdDSA) with a single message sent to the exchange server.

\section{Evaluation of ECDSA Threshold Signature Schemes}
\label{sec:evaluation}
\begin{table*}[!htb]
	\small
	\caption{Summary of ECDSA threshold signature schemes candidates}
	\centering
	\label{fig:evaluation}
	\begin{tabular}{l c c r r}
		\toprule
		\bfseries Scheme & \bfseries Threshold-optimal & \bfseries Rounds (signing) & \bfseries Signing time & \bfseries Signing bandwidth\\
		\midrule\midrule
		\citeauthor{lindell2017}~\cite{lindell2017}~\cite{cryptoeprint:2017:552} & \cmark & \num{2}  & $\sim$ \SI{35}{\milli\second} & $\sim$ \SI{0.8}{\kilo\byte} \\
		\citeauthor{8418649}~\cite{8418649} & \cmark & \num{2} & $\sim$ \SI{3}{\milli\second} & $\sim$ \SI{85.7}{\kilo\byte} \\
		\citeauthor{10.1007/978-3-030-26954-8_7}~\cite{10.1007/978-3-030-26954-8_7} & \cmark & \num{2} & $\sim$ \SI{150}{\milli\second} & $\sim$ \SI{5.6}{\kilo\byte} \\
		\bottomrule
	\end{tabular}
\end{table*}
We evaluate all recently proposed threshold ECDSA signature schemes (roughly in chronological order) with regard to their suitability to build the foundation of the protocol presented in this paper.

\citeauthor{cryptoeprint:2015:1169}~\cite{cryptoeprint:2015:1169} improved upon original work from~\citeauthor{1562272}~\cite{1562272} and~\citeauthor{10.1007/3-540-68339-9_31}~\cite{10.1007/3-540-68339-9_31}.
The authors introduced fully distributed key generation and rekeying and modified the signing algorithm slightly to improve performance.
However, the number of participants required is $\ge 2 \cdot t$, with $t$ being the threshold required to generate signatures.
Since neither the client nor the exchange must be able to generate signatures unilaterally, a threshold of $2$ means that we would need to introduce additional parties, which is sub-optimal in the setting of decentralized cryptocurrency exchanges.

\citeauthor{Gennaro2016Threshold}~\cite{Gennaro2016Threshold} published the first threshold-optimal scheme (for DSA/ECDSA), with $n > t$.
Signature generation requires \num{6} rounds of communication, however, which impairs fast execution.

\citeauthor{lindell2017} proposed a $2$-party ($2$-of-$2$) threshold signature scheme~\cite{lindell2017}~\cite{cryptoeprint:2017:552}\footnote{Note that the full version of the paper~\cite{cryptoeprint:2017:552} has received significant updates after the original publication.}.
While a \num{2}-party threshold signature scheme seems like a significant limitation at the first glance, it is sufficient for the protocol that encompasses a client and an exchange. 
The scheme requires just \num{2} rounds of communication for generating a signature, is bandwidth-efficient, and computationally relatively fast.
This efficiency is achieved by one party holding a homomorphic encryption of the other party’s key share.
This threshold ECDSA signature scheme is the first reasonable candidate to build the foundation of our protocol.

\citeauthor{boneh2017using}~\cite{boneh2017using} improve upon \citeauthor{Gennaro2016Threshold}'s work~\cite{Gennaro2016Threshold}, reducing the number of communication rounds required for signature generation from \num{6} to \num{4}, which is still more than the \num{2} rounds of~\cite{lindell2017}, and the back-and-forth communication creates additional latency that impairs efficient trading.

The $2$-of-$2$ and $2$-of-$n$ ECDSA threshold schemes proposed by \citeauthor{8418649}~\cite{8418649} do not rely on homomorphic encryption but instead use Oblivious Transfer (OT) multiplication.
Signature generation is extremely performant and requires just \num{2} rounds of communication, but the bandwidth requirements are rather high.
Nonetheless, it is a reasonable candidate.

\citeauthor{10.1145/3243734.3243788}~\cite{10.1145/3243734.3243788} published a full threshold ($t$-of-$n$) ECDSA signature scheme.
Instead of homomorphic encryption, the authors use ElGamal \enquote{in the exponent} to facilitate practical key generation in a multi-party setting and make signature generation more efficient.
For signature generation, however, \num{8} rounds of communication are needed, which makes the scheme impractical in our protocol.

\citeauthor{10.1145/3243734.3243859}~\cite{10.1145/3243734.3243859} improve upon \citeauthor{boneh2017using}'s work~\cite{boneh2017using}, reducing communication overhead and the time to generate signatures.
The number of communication rounds is increased to \num{9}, however, which creates additional latency that impairs efficient trading.

\citeauthor{cryptoeprint:2019:523}~\cite{cryptoeprint:2019:523}~\cite{8835354} improve upon their previous work~\cite{8418649}, extending it from a $2$-of-$n$ to a full $t$-of-$n$ threshold signature scheme.
The number of communication rounds required is $\log(t) + 6$, however, which makes the scheme impractical in our protocol.

\citeauthor{10.1007/978-3-030-26954-8_7}~\cite{10.1007/978-3-030-26954-8_7} generalized \citeauthor{lindell2017}'s $2$-of-$2$ threshold signature scheme~\cite{lindell2017} using hash proof systems.
The scheme achieves good performance and requires low bandwidth, which makes it the third reasonable candidate to build the foundation of API keys for our protocol.

\citeauthor{castagnos2020bandwidth}~\cite{castagnos2020bandwidth} build upon \citeauthor{10.1145/3243734.3243859}'s $t$-of-$n$ threshold signature scheme~\cite{10.1145/3243734.3243859}, improving computational and bandwidth efficiency.
The number of communication rounds required for generating signatures is reduced from \num{9} to \num{8}, which leaves the scheme still impractical for our use.

\subsection*{Summary of Evaluation}
While most recent $t$-of-$n$ ECDSA threshold signature schemes are threshold-optimal, our evaluation revealed that all of them require too many rounds of communication for signature generation.
$2$-of-$2$ or $2$-of-$n$ ECDSA threshold signature schemes, on the other hand, require only \num{2} rounds of communication, which significantly reduces latency and therefore facilitates efficient trading.
It remains an open question whether the higher number of rounds of $t$-of-$n$ ECDSA threshold signature schemes is a fundamental limitation or just requires further research.
In any case, for our protocol, a $2$-of-$2$ ECDSA threshold signature scheme is sufficient, and we continue our evaluation of the three candidate schemes.

Our evaluation resulted in three candidate schemes by \citeauthor{lindell2017}~\cite{lindell2017}, \citeauthor{8418649}~\cite{8418649}, and \citeauthor{10.1007/978-3-030-26954-8_7}~\cite{10.1007/978-3-030-26954-8_7}.
All three candidates are threshold-optimal and require just \num{2} rounds of communication for signature generation.
Apart from the cryptographic assumptions the schemes rely on, they differ in performance in terms of time required to generate signatures and the bandwidth required.
\citeauthor{lindell2017}'s scheme~\cite{lindell2017} has good performance in terms of time required to generate a signature ($\sim$\SI{35}{\milli\second}) and requires the least bandwidth ($\sim$\SI{0.8}{\kilo\byte}).
While \citeauthor{8418649}'s scheme~\cite{8418649} is the fastest, it also requires significantly more bandwidth.
Traders conducting thousands of trades per second would require tens of \si{\giga\bit} of bandwidth solely for generating signatures, which represents a significant disadvantage for \citeauthor{8418649}'s scheme.
\citeauthor{castagnos2020bandwidth}'s scheme ~\cite{castagnos2020bandwidth} is second of the three with regard to bandwidth required ($\sim$\SI{5.6}{\kilo\byte}) but brings a significant disadvantage in terms of time to generate a signature ($\sim$\SI{150}{\milli\second}), which makes it less favorable.
So \citeauthor{lindell2017}'s scheme~\cite{lindell2017} remains as the best contender overall --- in terms of bandwidth and in terms of time to generate a signature.
Table~\vref{fig:evaluation} summarizes the results.
And, as we will explain in Section~\ref{sec:protocol}, we improve the perceived signing time such that it is almost en par with \citeauthor{8418649}'s scheme~\cite{8418649}.

\section{API Keys for Decentralized Exchanges}
\label{sec:protocol}
\subsection{ECDSA Threshold Signature Schemes}
With a $2$-of-$2$ threshold signature scheme, the two parties, i.e., the user and the exchange, are both required to cooperate during day-to-day operations, i.e., generate signatures on transactions.
To this end, the full secret key $x$ is split into two multiplicative secret shares $x_1$ and $x_2$ where $x \equiv x_1 \cdot x_2 \mod q$, with $q$ being the order of the elliptic curve.
One key share remains with the user's client, i.e., the machine holding the API key, and the other key share goes to the exchange.
Signatures that would be valid under the public key cannot be generated with one key share alone and neither does knowledge of any one key share (i.e., $x_1$ or $x_2$) leak information about the full secret key $x$.
Both parties can therefore control user’s funds only when collaborating.
In this way, the exchange can restrict an API key's capability (by refusing to collaborate) without requiring full access to the user's funds.
Generally speaking, two parties interact in secret-key-operations (i.e., signing), and neither party can manipulate that operation to gain an advantage, even if that party is malicious.
The most harmful thing a malicious party can do during signature generation is to prevent the operation from completing.

For the reasons outlined in Section~\ref{sec:evaluation}, we build upon the signing algorithm from \citeauthor{lindell2017}'s scheme~\cite{lindell2017}.
However, there exist (significant) differences:
\begin{enumerate}
	\item API keys are generated solely by the client.
	\item We significantly improve performance by using Diffie-Hellman key exchange instead of a coin tossing protocol.
	\item We dramatically improve the perceived performance by employing a \num{2}-phase approach for generating signatures.
\end{enumerate}

\subsection{API Key Generation}
\label{sec:keygen}
An API key is generated by the user's machine with access to the full secret key.
Since that machine is considered a \enquote{trusted dealer}, API key generation can happen entirely on the client.
At first glance, this may sound counter-intuitive, but it is a result of the fact that we exploit knowledge of the full secret key on the client.
With this trick, the client can create API keys efficiently.

An API key consists of two secret shares: the client secret share and the (encrypted) exchange secret share.
The client sets the exchange's secret share $x_1$ initially to $1$ and the client's secret share $x_2$ to the full secret key $x$.
Then, a random number $r$ (not to be confused with the $r$-part of an ECDSA signature) is generated by the client and the secret shares are updated as follows:
\begin{enumerate}
	\item The update to the exchange's secret share is computed $x_1\prime \equiv x_1 \cdot r \mod q$.
	\item The new exchange's secret share $x_1\prime$ is encrypted under the exchange's Paillier public key.
	\item The update to the client's secret share is computed $x_2\prime \equiv x_2 \cdot r^{-1} \mod q$.
\end{enumerate}
Note that $x \equiv x_1 \cdot x_2 \equiv x_1\prime \cdot x_2\prime \mod q$, and therefore the full secret key $x$, the public key, and the corresponding address remain unchanged such that -- with the help of the API key -- signatures can be generated that are indistinguishable from signatures generated by the full secret key and are therefore oblivious to blockchains.

The client using the API key must not be able to gain knowledge of the full secret key.
For this reason, the exchange's secret share is encrypted under the exchange's Paillier public key.
Note that the client must verify that the Paillier public key was generated correctly.
A Paillier public key is defined as $N = p \cdot q$, with $p$, $q$ being large primes.
In order to certify that the exchange has generated the Paillier public key correctly, it needs to proof that $gcd(N, \phi(N)) = 1$ holds.
To this end, we use the proof from \citeauthor{GoldbergRSB19}~\cite{GoldbergRSB19} Section~3.2 with parameters $\alpha = 6370$ and $m_1 = m_2 = 11$ as suggested by \citeauthor{10.1145/3243734.3243788}~\cite{10.1145/3243734.3243788} in Section~6.3.2.
Since the exchange needs to create just a single Paillier key pair for all users, it needs to compute the proof of correctness just once.
On the client side, verifying the correctness of the Paillier public key needs to be conducted only once as well, but is computationally rather efficient ($\sim$\SI{41}{\milli\second}) anyway.

\subsection{Signature Generation: Preparation Phase}
For signature generation, we employ a \num{2}-phase approach.
The main idea with such a \num{2}-phase approach is to shift some required communication and computation into a message-independent preparation phase such that the finalization phase can happen with a single message from the client to the exchange.
By leveraging computational resources when a trader is not trading, performance is improved significantly.

With just one round of communication, client and server prepare a pool of (arbitrarily many) elliptic curve points, from which the first part of the signature is derived (as described in Section~\ref{sec:ecdsa}).
Recall that a point is computed as $R = k \cdot G$ where $k$ is a cryptographically secure nonce.
Knowledge of $k$ allows deriving the secret key from a signature, so neither client nor exchange must know $k$.
We employ the elliptic-curve Diffie-Hellman (ECDH) key agreement protocol in a bit of an unusual way.
Commonly, ECDH is used to generate a shared secret over an insecure channel such that an observer of the communication cannot derive the shared secret.
In our case, we employ ECDH such that the communicating parties arrive at a shared value (which will be made public afterwards) but do not know the other party's secret value.
In the threshold ECDSA setting $k \equiv k_1 \cdot k_2 \mod q$, and each party knows $R$ and either $k_1$ or $k_2$ but cannot derive $k$ (unless that party could solve the elliptic curve discrete logarithm problem, which is supposedly intractable).

The main advantage of this \num{2}-phase approach is that the signature finalization phase can be done computationally efficiently with just a single message from the client to the exchange.
This efficiency is achieved by shifting most of the communication and computational work into the signature preparation phase.
The computational work entails the (precomputed) randomness for Paillier encryption, which comprises modular exponentiation on big integers, and scalar multiplication.
The storage overhead is negligibly small with \SI{65}{\byte} per prepared point on the server (\SI{33}{\byte} for the compressed point representation and \SI{32}{\byte} for the server's DH secret)\footnote{Note that the provided numbers represent the minimum and may be slightly higher depending on the encoding used.}.
Storage overhead is also small on the client with \SI{321}{\byte} per prepared point on the client (\SI{33}{\byte} for the compressed point representation, \SI{32}{\byte} for the client's DH secret, and \SI{256}{\byte} for the Paillier randomness).
Computational cost and bandwidth required for the preparation phase increases linearly with the number of points to prepare, but this preparational step can happen anytime before the user wants to trade (e.g., right after login) such that it does not affect the performance of creating the final signature (which is effectively what traders perceive as latency).
Table~\vref{fig:signing_preparation} depicts the preparation phase of signature generation.

\begin{table}[!htb]
	\small
	\caption{Signature generation protocol: preparation phase}
	\centering
	\begin{threeparttable}
	\label{fig:signing_preparation}
	\begin{tabular}{l c l}
		\toprule
		\multicolumn{1}{c}{\bfseries Exchange} & & \multicolumn{1}{c}{\bfseries Client}\\
		\midrule\midrule
		& & Choose random $k_2$\\
		& & Compute $R_2 = k_2 \cdot G$\\
		& $\xleftarrow{R_2}$ &\\
		Choose random $k_1$ & &\\
		Compute $R_1 = k_1 \cdot G$ & &\\
		Compute $R = k_1 \cdot R_2$ & &\\
		Store $R$, $k_1$ & &\\
		& $\xrightarrow{R_1}$ &\\
		& & Compute $R = k_2 \cdot R_1$\\
		& & Compute random $\rho$\\
		& & Store $R$, $k_2$, $\rho$\\
		\bottomrule
	\end{tabular}
	\begin{tablenotes}
	\small
	\item $\rho$: (precomputed) randomness for the homomorphic Paillier encryption scheme.
	\item $G$: generator of the curve.
	\end{tablenotes}
	\end{threeparttable}
\end{table}

\subsection{Signature Generation: Finalization Phase}
\begin{table*}[!htb]
	\small
	\caption{Signature generation protocol: finalization phase}
	\centering
	\begin{threeparttable}
	\label{fig:signing_finalization}
	\begin{tabular}{l c l}
		\toprule
		\multicolumn{1}{c}{\bfseries Exchange} & & \multicolumn{1}{c}{\bfseries Client}\\
		\midrule\midrule
		& & Compute $r$ from $R$\\
		& & $c_3 = Enc(k_2^{-1} \cdot r \cdot d_2 \cdot d_{1enc} + k_2^{-1} \cdot m + \rho \cdot q)$\\
		& & Delete $R$, $k_2$, and $\rho$\\
		& $\xleftarrow{c_3, R}$ &\\
		Compute $r$ from $R$ & &\\
		Lookup $k_1$ with $R$ & &\\
		Compute $s = k_1^{-1} \cdot Dec(c3) \mod q$ & &\\
		Delete $R$ and $k_1$ & &\\
		Verify signature (r, s) & &\\
		\bottomrule
	\end{tabular}
	\begin{tablenotes}
	\small
	\item $d_2$: the client's secret share.
	\item $d_{1enc}$: the encrypted secret share of the exchange.
	\item $\rho$: the (precomputed) randomness for the Paillier homomorphic encryption scheme.
	\item $m$: the (message's) hash to be signed.
	\item $q$: the order of the curve.
	\end{tablenotes}
	\end{threeparttable}
\end{table*}

The signature finalization phase consists just of a single message from the client to the exchange, like with a centralized exchange where the client sends a single message indicating a trade.
In this way, there is no effectively communication overhead in terms of communication rounds.

The client chooses any of the precomputed points and computes a \textit{presignature} using the API key and that point.
The client then sends the presignature along with the chosen point over to the exchange to complete the signature.
It is essential for security that the client deletes the point used to prevent reuse (and therefore potential compromise of the secret key).
Upon ensuring compliance with the signing policy (see Section~\ref{sec:policy}), the exchange completes the signature.
Again, deletion of the point and DH secret used is paramount for security.
Eventually, the correctness of the signature is verified.
Table~\ref{fig:signing_finalization} depicts the finalization phase of signature generation.

\subsection{Performance}
With the \num{2}-phase approach described, no communication and computation is spared as such, but most communication and computation is shifted to a point in time when it does not impair trading.
In this way, trading can happen with a single message from the client to the exchange with relatively little computation required --- the performance perceived by traders when conducting trades on a decentralized exchange is now comparable to trading on a centralized exchange.

\begin{table}[!htb]
	\small
	\caption{Performance: preparation phase}
	\centering
	\label{fig:timing_preparation}
	\begin{tabular}{l c l}
		\toprule
		\multicolumn{1}{c}{\bfseries Exchange} & & \multicolumn{1}{c}{\bfseries Client}\\
		\midrule\midrule
		& & Time: $\sim$\SI{13}{\milli\second}\\
		& $\xleftarrow{\SI{33}{\byte}}$ &\\
		Time: $\sim$\SI{0.16}{\milli\second} & &\\
		& $\xrightarrow{\SI{33}{\byte}}$ &\\
		\bottomrule
	\end{tabular}
\end{table}

\begin{table}[!htb]
	\small
	\caption{Performance: finalization phase}
	\centering
	\label{fig:timing_finalization}
	\begin{tabular}{l c l}
		\toprule
		\multicolumn{1}{c}{\bfseries Exchange} & & \multicolumn{1}{c}{\bfseries Client}\\
		\midrule\midrule
		& & Time: $\sim$\SI{1.72}{\milli\second}\\
		& $\xleftarrow{\SI{545}{\byte}}$ &\\
		Time: $\sim$\SI{2.46}{\milli\second} & &\\
		\bottomrule
	\end{tabular}
\end{table}

Tables~\ref{fig:timing_preparation} and~\ref{fig:timing_finalization} show the performance for the preparation and finalization phases respectively.
With the \num{2}-phase approach, most of the communication and computation is moved into the preparation phase.
The preparation phase takes few computational and bandwidth resources, and the overall time is dominated by communication delays.
The finalization phase is very fast and takes just \SI{4}{\milli\second} in total and a single message from the client to the exchange\footnote{Note that the numbers depend on the elliptic curve used. For the measurements we used \textit{secp256k1}.}

\subsection{Signing Policy}
\label{sec:policy}
On a decentralized exchange, a full client without API keys signs a transaction and sends the signature to the exchange.
For API keys, however, the exchange receives a presignature from the client and needs to complete it first.
The exchange may base its agreement to complete a signature upon several factors, without taking custody of the user's funds.
These factors are specified as \textit{signing policy} by the user.
Obviously, not the holder of the API key but only the user must be able to create and modify the signing policy.
Such policies basically restrict the conditions under which the exchange participates in signature generation.
Policies may involve for example:
\begin{itemize}
	\item Daily / weekly / monthly withdrawal limits. The user can set an amount that is allowed to be moved from their exchange account each day. This can limit damage in the event of a compromised API key.
	\item IP address restrictions for trading and/or withdrawal.
	\item Location restrictions for trading and/or withdrawal.
	\item Withdrawal addresses restrictions.
	\item Daily / weekly / monthly trading limits. Limits are imposed upon how much users can trade.
	\item Time delayed withdrawals, where funds will be moved only after a \num{24}- or \num{48}-hour period. Such delay allow users in the event of a hacked account to reach out to the exchange and prevent fund movement within the delay period.
	\item Allow access only from specific devices.
	\item Allow access only to specific markets.
	\item Restrictions based on biometric information such as a fingerprint scan.
\end{itemize}



\section{Discussion and Future Work}
\begin{table*}[!htb]
	\small
	\caption{Paillier performance}
	\centering
	\label{fig:paillier}
	\begin{tabular}{l r r r r r r}
		\toprule
		\bfseries Operation / Key size & \bfseries \SI{512}{\bit} & \bfseries \SI{1024}{\bit} & \bfseries \SI{2048}{\bit} & \bfseries \SI{3072}{\bit} & \bfseries \SI{4096}{\bit} & \bfseries \SI{8192}{\bit}\\
		\midrule\midrule
		Precompute randomness & \SI{320}{\micro\second} & \SI{2106}{\micro\second} & \SI{13595}{\micro\second} & \SI{39583}{\micro\second} & \SI{74151}{\micro\second} & \SI{420651}{\micro\second}\\
		Encrypt & $<$\SI{1}{\micro\second} & \SI{1}{\micro\second} & \SI{4}{\micro\second} & \SI{7}{\micro\second} & \SI{12}{\micro\second} & \SI{35}{\micro\second}\\
		Decrypt & \SI{106}{\micro\second} & \SI{453}{\micro\second} & \SI{2081}{\micro\second} & \SI{6032}{\micro\second} & \SI{13666}{\micro\second} & \SI{78387}{\micro\second}\\
		\bottomrule
	\end{tabular}
\end{table*}

\subsection{Threshold EdDSA}
We have extended the API key protocol described in Section~\ref{sec:protocol} to also work with the EdDSA signature scheme.
In contrast to threshold ECDSA (where multiplicative secret sharing is used), we use additive secret sharing for threshold EdDSA.
For API key generation, the client generates a random scalar $r$ which represents its secret share and computes the server's secret share as $a - r \mod L$, where $a$ is the signing key.
The client keeps its secret share and sends the (encrypted) exchange secret share to the exchange.

For signature generation we employ the same DH-like approach as we do for threshold ECDSA.
The resulting point is computed a bit different though as $R$ is the (point) addition of the public client point plus the public server point.
While using that approach breaks EdDSA’s deterministic approach, it does generates valid signatures nonetheless.
Like in the threshold ECDSA scheme, we use the two-phased approach with a message-independent phase and a second phase in which the signature for a particular message is generated in order to facilitate signature generation with a single message from client to server and therefore improve the performance perceived by users.

Signature generation for threshold EdDSA is straight-forward: the client can compute its part of the second part of the signature $s$ as ${s\_client} = {r\_client} + h * {client\_secret\_share}$, with $h = H(r | A | M)$ ($r$: first part of the signature, $A$: public key, $M$: message).
${s\_client}$ is sent over to the server.
The server computes its corresponding part and adds both numbers, which results in the second part of the signature $s$.
The computations are very efficient, since the client does not need to do computations on the encrypted server secret share.
For this reason, the computational performance of signature generation in threshold EdDSA is exceptional - just marginally slower than conventional, non-threshold EdDSA.

\subsection{Generalization to Other Decentralized Services}
In this paper, we explored the application of decentralized cryptocurrency exchanges, i.e., the Nash exchange.
The protocol we presented, however, can be generalized to other decentralized services, and we have already applied the protocol to Uniswap and 1inch, allowing users to trade via the associated smart contracts without accessing their full secret key.

\subsection{Paillier Key Size}
The currently recommended key size for the Paillier cryptosystem is \SI{2048}{\bit}.
Increasing the key size would be desirable but larger keys significantly increase the times for encrypting and decrypting.
Table~\vref{fig:paillier} shows the results of our evaluation.
Precomputing randomness takes longest, but is part of the signature preparation phase and does not affect signature finalization.
Encryption increases as well with larger key sizes but the time it takes is negligible even with very large keys.
Decryption, however, takes a significant amount of time and is part of the signature finalization phase.
We leave the question of how to optimize the performance of Paillier decryption with large keys as future work.


\section{Conclusion}
In this paper, we presented a protocol enabling API keys for decentralized cryptocurrency exchanges that is suitable for both security-sensitive end-users and professional traders alike.
Given the fact that the full secret key is never required for any day-to-day operation, the risk of compromised secret keys is drastically reduced.
The protocol presented in this paper allows users of an exchange to create many API keys associated with their account, each of which may be entitled to different (withdrawal and trading) limits.
This allows trading companies to use decentralized exchanges without trusting their employees with credentials that have access to the full secret key material.
To this end, client and exchange engage in an interactive signature generation protocol.
The communication overhead is negligible, and the effective delay due to computations is roughly \SI{4}{\milli\second}, satisfying the low latency requirements of professional traders.
High performance is achieved by employing a \num{2}-phase approach, which shifts the majority of the computational time required as well as the additional communication delay (caused by the interactivity of the protocol) into a message-independent phase.
In this way, a signature can be finalized with a single message from client to the exchange.
We additionally provide an open source implementation of the protocol in Rust~\footnote{\url{https://github.com/nash-io/nash-rust/tree/master/mpc-wallet/nash-mpc}}.

\balance

\printbibliography[heading=bibintoc]

\end{document}